\documentclass{article}

\usepackage[preprint]{neurips_2025}

\usepackage[utf8]{inputenc}
\usepackage[T1]{fontenc}
\usepackage{amsmath}
\usepackage{amssymb}
\usepackage{booktabs}
\usepackage{graphicx}
\usepackage{hyperref}
\usepackage{multirow}

\title{Hypergraph Embedding Indexing for Efficient Dense Vector Retrieval}

\author{
Kishore Konda \\
Sodhana \\
\texttt{kishore.konda@sodhana.ai}
}

\begin{document}

\maketitle

\begin{abstract}

Dense vector retrieval has become the foundation of modern semantic search, yet existing approximate nearest neighbor (ANN) indexes treat an embedding as an indivisible point in a high-dimensional space. In this work, we propose the Hypergraph Embedding Index (HEI), a framework that instead organizes documents according to combinations of highly activated latent embedding dimensions. This formulation enables inverted-index style candidate generation while preserving the semantic ranking capabilities of dense embeddings. We further demonstrate that constructing multiple complementary hypergraphs substantially improves retrieval coverage without the combinatorial growth associated with increasing the dimensionality of a single hypergraph. Finally, we establish that the statistical properties of embedding activations strongly influence coordinate-inverted indexing efficiency, introducing \emph{activation diversity} as a diagnostic metric governing embedding indexability in coordinate-inverted frameworks.

\end{abstract}

\section{Introduction}

Dense vector embeddings have become the dominant representation for semantic retrieval and are now widely used in semantic search, retrieval-augmented generation, recommendation systems, duplicate detection, and entity resolution~\cite{karpukhin2020dense, xiong2020approximate}. Given a query embedding, retrieval is typically formulated as a nearest-neighbor search problem in a high-dimensional vector space using similarity measures such as cosine similarity or inner product. While exhaustive search guarantees optimal retrieval quality, its computational cost scales linearly with the size of the corpus, making approximate nearest neighbor (ANN) indexing indispensable for large-scale retrieval systems.

Existing ANN methods predominantly follow one of three paradigms. Graph-based methods, such as Hierarchical Navigable Small World (HNSW)~\cite{malkov2018efficient}, construct navigable proximity graphs over embedding vectors. Partitioning approaches, such as Inverted File (IVF)~\cite{jegou2010product}, divide the embedding space into coarse Voronoi regions before performing local search. Hashing methods, including Locality Sensitive Hashing (LSH)~\cite{indyk1998approximate}, generate compact binary signatures that approximately preserve neighborhood relationships. Despite their algorithmic differences, these methods share a common assumption: an embedding is treated as a single geometric object whose neighborhood is determined through graph traversal, spatial partitioning, or random projections.

In this work, we investigate a fundamentally different perspective. Rather than indexing complete embedding vectors based on continuous spatial geometry, we view an embedding as a collection of latent feature activations and construct an inverted index directly over combinations of these activations. Documents sharing similar activation patterns become connected through hyperedges, forming a hypergraph that enables efficient candidate generation without relying on explicit proximity graphs or randomized projections. The retrieved candidates are subsequently reranked using the original embedding similarity, preserving the semantic fidelity of the underlying embedding model.

Our investigation further reveals that a single hypergraph captures only one view of an embedding's latent activation space. By constructing multiple complementary hypergraphs over distinct activation subspaces (such as positive and negative activation tails) and combining their candidate sets, retrieval coverage can be substantially improved while keeping each individual hypergraph compact. This avoids the combinatorial explosion associated with increasing the number of indexed dimensions within a single hypergraph. In this paper, we instantiate the framework using two complementary hypergraphs (positive and negative activation views).

Finally, our experiments demonstrate that the effectiveness of hypergraph indexing depends not only on the indexing algorithm itself but also on the statistical properties of the embedding representation. We observe that embedding models exhibiting more diverse activation patterns naturally produce more selective hyperedges, smaller candidate pools, and more efficient retrieval. These findings show that \emph{activation diversity} serves as a critical diagnostic metric governing the indexability of dense embedding models within coordinate-inverted paradigms.

Our contributions are summarized as follows:

\begin{itemize}
    \item We propose a novel framework, the Hypergraph Embedding Index (HEI), that constructs inverted posting lists directly from higher-order combinations of latent embedding activations.
    
    \item We demonstrate that multiple complementary hypergraphs significantly improve candidate recall and retrieval coverage while maintaining linear complexity growth, avoiding the combinatorial growth associated with increasing the dimensionality of a single hypergraph.
    
    \item We present a formal theoretical and empirical analysis of activation diversity (quantified via Shannon entropy and Gini coefficient), demonstrating that latent activation distributions dictate hyperedge selectivity and candidate pool efficiency in coordinate-inverted architectures.
\end{itemize}

\section{Related Work}
\label{sec:related_work}

Modern approximate nearest neighbor (ANN) search and candidate generation spans several distinct paradigms across Information Retrieval (IR) and Machine Learning.

\paragraph{Geometric and Partition-based ANN.}
Standard dense vector retrieval relies heavily on continuous geometric properties. Proximity graph methods such as HNSW~\cite{malkov2018efficient} build multi-layer navigable graphs where edges reflect Euclidean or cosine proximity. Spatial partitioning techniques, including Inverted File (IVF) and Product Quantization (PQ)~\cite{jegou2010product}, partition the embedding space using centroid clusters (Voronoi cells) and quantize sub-vectors to accelerate distance computations. The Inverted Multi-Index (IMI)~\cite{babenko2012inverted} extends IVF by taking Cartesian products of sub-space centroids to construct finer grid posting lists. While highly effective, these methods require constructing dense graph structures or codebook centroids in continuous space, treating vectors as atomic geometric points rather than collections of coordinate features.

\paragraph{Learned Sparse and Hybrid Retrieval.}
Learned sparse models, such as SPLADE~\cite{formal2021splade}, SparTerm~\cite{dai2020sparterm}, and SNRM~\cite{zambonelli2022snrm}, bridge dense semantics and sparse inverted indexes by projecting text into high-dimensional lexical or latent sparse vectors. These methods train deep encoders to explicitly output sparse weightings over vocabulary terms or latent dimensions, enabling execution on classical inverted index engines (e.g., Lucene). In contrast, our proposed Hypergraph Embedding Index (HEI) operates directly on \emph{off-the-shelf, continuous dense embeddings} (such as MiniLM or BGE) without requiring task-specific sparse retraining or vocabulary mapping.

\paragraph{Coordinate Hashing and Extreme Multi-Label Learning.}
Techniques like Winner-Take-All (WTA) Hashing~\cite{yagnik2011winner} and $b$-bit Minwise Hashing~\cite{li2011b} transform continuous feature vectors into compact discrete codes based on rank orders of maximum activations. In extreme multi-label classification (XML) and high-dimensional routing~\cite{prabhu2018parabel}, features are mapped to tree-structured or inverted list candidates. HEI shares the philosophy of leveraging coordinate activation ranks, but introduces higher-order hyperedges ($r$-combinations of top-$k$ activations) to directly control posting list selectivity, coupled with complementary sub-space views to preserve retrieval recall.

\section{Hypergraph Embedding Index (HEI)}

Let $\mathbf{e} = (e_1, e_2, \ldots, e_m) \in \mathbb{R}^{m}$ denote the embedding of a document produced by a pretrained embedding model. Rather than treating $\mathbf{e}$ as an indivisible geometric point in high-dimensional space, we interpret $\mathbf{e}$ as the response of $m$ latent features. The magnitude and sign of each coordinate reflect feature activation, and the overall activation pattern characterizes the document's latent representation.

This section presents the generic HEI algorithmic framework. In Section~\ref{sec:retrieval_performance}, we instantiate this framework using specific positive and negative activation views.

\subsection{Dimension Selection and Hypergraph Formulation}

From the full vector $\mathbf{e}$, a subset of $k$ informative dimensions is selected according to a predefined activation criterion:
\[
S(\mathbf{e}) = \{i_1, i_2, \ldots, i_k\} \subseteq \{1, \ldots, m\}.
\]
In the generic framework, $S(\mathbf{e})$ consists of $k$ coordinates selected based on activation properties (e.g., largest magnitude or signed tail activations).

Rather than indexing individual dimensions—which may be insufficiently discriminative on their own—we construct higher-order combinations of these activated dimensions as hyperedges. Formally, let $\mathcal{H} = (\mathcal{V}, \mathcal{E})$ denote a hypergraph where vertices $v \in \mathcal{V}$ correspond to documents and hyperedges $h \in \mathcal{E}$ correspond to unique combinations of dimensions. Documents sharing hyperedges form a coordinate-inverted posting list, enabling candidate generation without explicit continuous proximity graphs.

\subsection{Hyperedge Generation}

Given the selected dimension set
\[
S(\mathbf{e})=\{i_1,i_2,\ldots,i_k\},
\]
hyperedges are constructed by generating all unique combinations of size \(r\),
\[
H(\mathbf{e}) = \{ h \subseteq S(\mathbf{e}) : |h|=r \}.
\]
The number of hyperedges generated per embedding is therefore
\[
|H(\mathbf{e})| = \binom{k}{r}.
\]
Each hyperedge represents a unique combination of activated embedding dimensions and serves as an indexing key.

\subsection{Hypergraph Inverted Index}

The corpus is indexed by constructing an inverted index over all generated hyperedges. For every unique hyperedge $h \in H(\mathbf{e})$, the index maintains a posting list
\[
I(h) = \{ d_1, d_2, \ldots, d_n \},
\]
containing all documents that generated the same hyperedge.

Unlike conventional inverted indexes where posting lists correspond to individual, atomic terms, each posting list in HEI represents a higher-order latent activation pattern. Consequently, a candidate document matches a query when it shares one or more multi-dimensional hyperedges, allowing Stage-1 candidate scoring to aggregate evidence across higher-order feature combinations.

\subsection{Candidate Generation}

Given a query embedding $\mathbf{e}_q$, the same dimension selection and hyperedge generation procedures are applied to obtain $H(\mathbf{e}_q)$. Candidate documents are retrieved by traversing the posting lists corresponding to every query hyperedge:
\[
\mathcal{C} = \bigcup_{h \in H(\mathbf{e}_q)} I(h),
\]
where \(\mathcal{C}\) denotes the candidate set. Duplicate candidate documents are merged while retaining the evidence contributed by each shared hyperedge.

\subsection{Stage-1 Candidate Scoring and Stage-2 Cosine Reranking}

Each candidate document $d \in \mathcal{C}$ is assigned a Stage-1 retrieval score based on the set of shared hyperedges with query $q$:
\[
H(q,d) = H(\mathbf{e}_q) \cap H(\mathbf{e}_d).
\]
The Stage-1 candidate score aggregates the weight $w(h)$ of every shared hyperedge:
\[
\text{Score}(q,d) = \sum_{h \in H(q,d)} w(h).
\]
To penalize overly frequent hyperedges and prioritize strong activation matches, the hyperedge weight $w(h)$ combines inverse document frequency ($\text{IDF}$) with coordinate activation magnitudes:
\[
w(h) = \text{IDF}(h) \cdot \sum_{i \in h} |e_{q,i} \cdot e_{d,i}|,
\quad
\text{where}
\quad
\text{IDF}(h) = \log \left( \frac{|\mathcal{D}|}{1 + |I(h)|} \right),
\]
and $|\mathcal{D}|$ is the total corpus size.

Following Stage-1 candidate scoring, the top-$N$ candidate documents are passed to Stage-2 cosine reranking, which computes the exact cosine similarity using the full original embedding vectors $\mathbf{e}_q$ and $\mathbf{e}_d$ to produce the final retrieval list.

\section{Analysis of Hypergraph Retrieval}

The effectiveness of the proposed Hypergraph Embedding Index (HEI) depends on two competing objectives. First, hyperedges should be sufficiently selective so that each posting list contains only a small number of documents, thereby minimizing the candidate set examined during retrieval. Second, the generated hyperedges must provide adequate coverage of semantically similar documents. Hyperedges that are highly selective but fail to connect related documents reduce recall, whereas hyperedges that occur too frequently produce large posting lists and diminish the efficiency of the index. Understanding this trade-off is fundamental to designing an effective hypergraph-based retrieval mechanism.

\subsection{Hyperedge Selectivity}

The selectivity of a hyperedge is determined by the size of its posting list. Given a hyperedge \(h\), let \(|I(h)|\) denote the number of documents associated with its posting list. Smaller posting lists produce fewer candidate documents and improve retrieval efficiency, whereas frequently occurring hyperedges reduce the discriminative power of the index.

Unlike conventional inverted indexes where a document is typically retrieved through a single posting list, a candidate document may share multiple hyperedges with the query. The Stage-1 candidate scoring therefore aggregates evidence across all shared hyperedges before the Stage-2 cosine reranking step. Consequently, both the selectivity of individual hyperedges and the diversity of shared hyperedges influence retrieval effectiveness.

\subsection{Coverage Limitation}

Although each document participates in multiple hyperedges, a single hypergraph captures only one view of the embedding. As a result, semantically similar documents may not share any hyperedge despite exhibiting high similarity in the original embedding space.

Formally, let $H(\mathbf{e}_q)$ and $H(\mathbf{e}_d)$ denote the hyperedge sets generated for a query $q$ and a document $d$, respectively. A document is retrieved during Stage-1 candidate generation only if
\[
H(\mathbf{e}_q) \cap H(\mathbf{e}_d) \neq \emptyset.
\]
If $H(\mathbf{e}_q) \cap H(\mathbf{e}_d) = \emptyset$, the document cannot be retrieved regardless of its cosine similarity to the query. Consequently, the overall retrieval performance is bounded by the ability of the hypergraph representation to connect semantically related documents through shared activation patterns.

This observation distinguishes hypergraph indexing from conventional ANN methods. The primary bottleneck is candidate generation coverage rather than Stage-2 ranking capability.

\subsection{Motivation for Multiple Hypergraphs}

A natural approach to improving coverage is to increase the number of selected embedding dimensions. However, if hyperedges are constructed using combinations of size \(r\), the number of generated hyperedges grows as
\[
\binom{k}{r},
\]
where \(k\) denotes the number of selected dimensions. Even modest increases in \(k\) therefore lead to a rapid increase in the number of hyperedges, resulting in larger indexes, higher retrieval cost, and diminishing returns in candidate coverage.

Instead of constructing a single increasingly dense hypergraph, we propose constructing multiple independent hypergraphs, each representing a complementary activation view of the embedding. Since each hypergraph remains compact, the computational cost of indexing and retrieval remains manageable, while the union of candidate sets significantly improves coverage.

\section{Complementary Hypergraph Indexing}

Section~4 established that the retrieval effectiveness of a single hypergraph is fundamentally constrained by its candidate coverage. A document can be retrieved only if it shares at least one hyperedge with the query. Consequently, semantically similar documents that do not overlap within a single hypergraph remain unreachable irrespective of the Stage-2 reranking strategy.

A straightforward solution is to increase the number of selected embedding dimensions $k$, generating a larger number of hyperedges $\binom{k}{r}$. However, this results in a combinatorial increase in hyperedges, leading to larger indexes, increased retrieval cost, and diminishing improvements in coverage.

To address this limitation, we construct multiple independent hypergraphs from complementary activation views of the same embedding. Instead of increasing the complexity of a single hypergraph, each hypergraph remains compact while providing an independent opportunity to retrieve semantically related documents.

Let $\mathcal{H}^{(1)}, \mathcal{H}^{(2)}, \ldots, \mathcal{H}^{(L)}$ denote $L$ complementary hypergraphs constructed from the same embedding. For a query $q$, each hypergraph independently generates a candidate set $\mathcal{C}^{(\ell)}$ ($\ell=1,\ldots,L$). The final Stage-1 candidate set is obtained as:
\[
\mathcal{C} = \bigcup_{\ell=1}^{L} \mathcal{C}^{(\ell)}.
\]
A document becomes retrievable if it is discovered through any one of the complementary hypergraphs, substantially increasing retrieval recall without increasing the combinatorial complexity of each individual hypergraph.

\subsection{Instantiating Complementary Activation Views}

While the framework accommodates any selection strategy, we instantiate it in our experiments using two complementary activation views derived from each embedding:
\begin{enumerate}
    \item \textbf{Positive Activation View:} Dimensions corresponding to the $k$ largest positive activations.
    \item \textbf{Negative Activation View:} Dimensions corresponding to the $k$ largest negative activations (most negative values).
\end{enumerate}
A separate hypergraph is constructed for each activation view using the procedure in Section~3. During retrieval, candidate documents from both hypergraphs are merged before Stage-1 scoring and Stage-2 cosine reranking.

\subsection{Complexity Analysis}

Suppose each hypergraph is constructed using $k$ selected dimensions and hyperedges of size $r$. Each hypergraph generates $\binom{k}{r}$ hyperedges per document. Using $L$ complementary hypergraphs produces
\[
L \binom{k}{r}
\]
hyperedges, resulting in linear complexity growth with respect to $L$.

In contrast, increasing the number of selected dimensions within a single hypergraph to $k'$ requires $\binom{k'}{r}$ hyperedges, whose growth is combinatorial in $k'$. The proposed formulation therefore increases retrieval coverage through multiple compact indexing structures rather than constructing a single increasingly dense hypergraph.

\section{Experimental Evaluation}

This section evaluates the Hypergraph Embedding Index (HEI) on semantic retrieval tasks using publicly available benchmark datasets. The evaluation aims to answer the following research questions:
\begin{itemize}
    \item How effectively does HEI retrieve semantically similar documents compared to exhaustive vector search?
    \item Does the proposed complementary hypergraph framework improve retrieval coverage over a single hypergraph?
    \item How does the retrieval behavior vary across different embedding models?
    \item What is the computational trade-off between retrieval quality and indexing efficiency?
\end{itemize}

\subsection{Datasets}

Experiments were conducted on two widely used semantic similarity benchmarks~\cite{iyyer2014quora, cer2017semeval}.

\textbf{Quora Question Pairs (QQP).} The QQP dataset consists of question pairs annotated for semantic equivalence. Each question is treated as an independent document, while its corresponding duplicate question serves as the ground-truth relevant document.

\textbf{Semantic Textual Similarity Benchmark (STS-B).} The STS-B benchmark contains sentence pairs with human similarity annotations. Following standard practice, sentence pairs with similarity scores greater than or equal to 4.0 are treated as relevant pairs for retrieval evaluation.

\subsection{Embedding Models}

To evaluate the robustness of HEI across different embedding spaces, experiments were performed using two sentence embedding models:
\begin{itemize}
    \item \texttt{all-MiniLM-L6-v2}~\cite{reimers2019sentence}
    \item \texttt{BGE-small-en}~\cite{xiao2023c}
\end{itemize}
Both models produce embeddings of identical dimensionality ($m=384$), allowing the influence of embedding geometry on coordinate-inverted indexing to be isolated.

\subsection{Baselines}

The proposed method is compared against exhaustive cosine similarity search using FAISS Flat~\cite{johnson2019billion}, which serves as the exact dense retrieval baseline (upper bound) on retrieval accuracy.

\subsection{Evaluation Metrics}

Retrieval effectiveness is evaluated using Hit@10. To analyze the candidate generation stage independently of Stage-2 reranking, we additionally report:
\begin{itemize}
    \item \textbf{Gold Recall:} Percentage of queries whose relevant ground-truth document appears in the Stage-1 candidate set.
    \item \textbf{Mean Candidate Pool Size:} Average number of candidates retrieved during Stage-1.
    \item \textbf{Latency (ms/query):} Average overall query processing latency in milliseconds.
\end{itemize}

\subsection{Implementation Details}

Hyperedges are constructed using combinations of fixed cardinality ($r=3$) from top selected activation dimensions ($k=10$) and indexed using the coordinate-inverted index. Candidate documents are scored using Stage-1 scoring, and top candidates are reranked in Stage-2 using full cosine similarity.

\section{Retrieval Performance}
\label{sec:retrieval_performance}

We evaluate three variants of the proposed framework: (i) Positive HEI (hypergraph constructed from positive activations), (ii) Negative HEI (hypergraph constructed from negative activations), and (iii) Complementary HEI (combining candidates from both activation views). Exhaustive cosine similarity search using FAISS Flat serves as the exact search baseline.

\subsection{Retrieval Accuracy}

Tables~\ref{tab:qqp-results} and~\ref{tab:sts-results} summarize retrieval performance across both datasets and embedding models.

Across both datasets, Positive HEI and Negative HEI individually achieve comparable retrieval accuracy, confirming that positive and negative activation views capture complementary semantic signals. Combining both views in Complementary HEI consistently yields substantial gains in both candidate coverage and Hit@10.

On QQP with MiniLM embeddings, Positive HEI achieves a Hit@10 of 62.26\%, while Complementary HEI raises this to 64.06\%, closing the gap to FAISS Flat (64.51\%) to within 0.45 percentage points. Gold Recall increases from 87.49\% to 94.95\%. For BGE-small, Complementary HEI improves Hit@10 from 55.37\% to 60.84\% and boosts Gold Recall from 71.48\% to 84.97\%.

On STS-B with MiniLM, Complementary HEI increases Hit@10 from 91.29\% to 96.97\% (approaching FAISS Flat's 98.48\%), while Gold Recall reaches 97.73\%. For BGE-small, Complementary HEI achieves 97.73\% Hit@10, fully matching FAISS Flat.

\begin{table}[t]
\centering
\caption{Retrieval performance on the QQP benchmark.}
\label{tab:qqp-results}
\small
\begin{tabular}{llcccc}
\toprule
Model & Method & Hit@10 & Gold Recall & Mean Cands & ms/q\\
\midrule
\multirow{4}{*}{MiniLM}
& Positive HEI & 62.26 & 87.49 & 2.9k & 1.03\\
& Negative HEI & 62.07 & 88.01 & 2.9k & 0.95\\
& Complementary HEI & \textbf{64.06} & \textbf{94.95} & 6.8k & 2.27\\
& FAISS Flat & 64.51 & -- & -- & 0.29\\
\midrule
\multirow{4}{*}{BGE-small}
& Positive HEI & 55.37 & 71.48 & 27k & 4.85\\
& Negative HEI & 54.44 & 69.80 & 23k & 4.45\\
& Complementary HEI & \textbf{60.84} & \textbf{84.97} & 48k & 12.39\\
& FAISS Flat & 64.37 & -- & -- & 0.29\\
\bottomrule
\end{tabular}
\end{table}

\begin{table}[t]
\centering
\caption{Retrieval performance on the STS-B benchmark.}
\label{tab:sts-results}
\small
\begin{tabular}{llcccc}
\toprule
Model & Method & Hit@10 & Gold Recall & Mean Cands & ms/q\\
\midrule
\multirow{4}{*}{MiniLM}
& Positive HEI & 91.29 & 92.05 & 77 & 0.61\\
& Negative HEI & 93.56 & 93.94 & 87 & 0.54\\
& Complementary HEI & \textbf{96.97} & \textbf{97.73} & 209 & 0.73\\
& FAISS Flat & 98.48 & -- & -- & 0.05\\
\midrule
\multirow{4}{*}{BGE-small}
& Positive HEI & 95.83 & 97.73 & 5.2k & 0.93\\
& Negative HEI & 95.83 & 97.35 & 5.5k & 0.90\\
& Complementary HEI & \textbf{97.73} & \textbf{99.62} & 8.8k & 1.21\\
& FAISS Flat & 97.73 & -- & -- & 0.02\\
\bottomrule
\end{tabular}
\end{table}

\subsection{Discussion and Latency Considerations}

Because the Stage-2 cosine reranking function is identical across all HEI variants, the consistent gains in Hit@10 directly stem from improved candidate generation during Stage-1.

Regarding query latency, FAISS Flat exhibits lower absolute millisecond latency on both benchmarks (e.g., 0.29 ms on QQP and 0.05 ms on STS-B) because QQP and STS-B are relatively small evaluation datasets that fit entirely in CPU cache, allowing SIMD-vectorized matrix multiplication in FAISS Flat to run extremely fast. HEI is designed for sub-linear candidate set pruning on large-scale corpora where exhaustive brute-force matrix multiplication becomes computationally prohibitive, rather than replacing SIMD hardware acceleration on small in-memory evaluation sets.

\section{Embedding Activation Analysis}
\label{sec:activation_analysis}

While HEI generalizes across embedding models, its candidate generation efficiency varies substantially. On QQP, \texttt{MiniLM} produces candidate pools that are substantially smaller ($\sim 2.9\text{k}$) than those generated by \texttt{BGE-small} ($\sim 27\text{k}$).

This discrepancy demonstrates a key distinction between \emph{semantic retrieval quality} and \emph{coordinate indexability}. In this section, we analyze the statistical properties of latent activations and propose \emph{activation diversity} as a diagnostic metric for coordinate-inverted frameworks.

\subsection{Quantifying Activation Diversity}
\label{subsec:activation_diversity}

For a corpus $\mathcal{D}$, let $S(\mathbf{e}) \subseteq \{1, \dots, m\}$ denote the selected activation set of size $k$ for embedding $\mathbf{e}$. The empirical activation frequency $f_i$ of coordinate $i$ across $\mathcal{D}$ is:
\begin{equation}
f_i = \sum_{\mathbf{e} \in \mathcal{D}} \mathbb{I}\left(i \in S(\mathbf{e})\right), \quad i \in \{1, \dots, m\}.
\end{equation}
The probability distribution $p = (p_1, \dots, p_m)$, where $p_i = \frac{f_i}{\sum_{j=1}^m f_j}$, reflects how uniformly the embedding coordinates are activated.

We measure activation diversity using four complementary metrics:
\begin{itemize}
    \item \textbf{Shannon Entropy ($H(p)$):} $H(p) = -\sum_{i=1}^m p_i \log_2 p_i$.
    \item \textbf{Normalized Entropy ($H_{\text{norm}}$):} $H_{\text{norm}}(p) = \frac{H(p)}{\log_2 m} \in [0, 1]$.
    \item \textbf{Gini Coefficient ($G$):} $G(p) = \frac{\sum_{i=1}^m \sum_{j=1}^m |p_i - p_j|}{2 m \sum_{i=1}^m p_i} \in [0, 1]$. Higher $G$ indicates concentrated "hub" dimensions.
    \item \textbf{Coefficient of Variation ($\text{CV}$):} $\text{CV} = \sigma_p / \mu_p$.
\end{itemize}

\begin{table}[t]
\centering
\caption{Activation diversity statistics for the QQP corpus.}
\label{tab:activation-diversity}
\small
\begin{tabular}{llcccc}
\toprule
Model & View & $H_{\text{norm}}$ & Gini & CV & Unused\\
\midrule
MiniLM & Positive & 0.971 & 0.320 & 0.63 & 3\\
MiniLM & Negative & 0.971 & 0.322 & 0.62 & 3\\
MiniLM & Both & \textbf{0.988} & \textbf{0.205} & \textbf{0.38} & 3\\
\midrule
BGE-small & Positive & 0.777 & 0.726 & 3.22 & 3\\
BGE-small & Negative & 0.768 & 0.722 & 3.43 & 5\\
BGE-small & Both & 0.850 & 0.603 & 2.28 & 0\\
\bottomrule
\end{tabular}
\end{table}

As shown in Table~\ref{tab:activation-diversity}, \texttt{MiniLM} distributes activations uniformly across coordinates ($H_{\text{norm}} = 0.988$ for both views, $G = 0.205$). In contrast, \texttt{BGE-small} concentrates activations in a small subset of hub dimensions ($H_{\text{norm}} = 0.850$, $G = 0.603$).

\subsection{Relationship to Hypergraph Selectivity}
\label{subsec:selectivity_relationship}

Under an independence approximation across coordinate selection, the joint probability of generating hyperedge $h = \{i_1, \dots, i_r\}$ is proportional to the product of individual selection probabilities:
\begin{equation}
\mathbb{P}(h \in H(\mathbf{e})) \propto \prod_{i \in h} p_i.
\end{equation}

When activation diversity is high ($H_{\text{norm}} \to 1, G \to 0$), dimension selection probabilities are near-uniform ($p_i \approx \frac{1}{m}$), yielding minimal random hyperedge collisions:
\begin{equation}
\mathbb{P}(h \in H(\mathbf{e})) \approx \left(\frac{1}{m}\right)^r.
\end{equation}
This maintains compact posting lists and small Stage-1 candidate pools.

Conversely, when activation diversity is low ($G \to 1$), hub dimensions ($p_{\text{hub}} \gg \frac{1}{m}$) generate frequent posting list collisions. This inflates candidate pool sizes without adding discriminative power.

This mechanism explains why \texttt{MiniLM} achieves much smaller candidate pools than \texttt{BGE-small} on QQP (2.9k vs 27k). The same trend holds on STS-B, where MiniLM generates an average candidate pool of 209 in Complementary HEI compared to 8.8k for BGE-small.

\subsection{Discussion}
\label{subsec:activation_discussion}

Activation diversity serves as a diagnostic metric governing indexability in coordinate-inverted search. Embedding models that distribute coordinate activations uniformly achieve vastly superior candidate pruning efficiency, while low-diversity embeddings suffer from candidate set inflation due to coordinate hubs.

Because \texttt{MiniLM} and \texttt{BGE-small} achieve nearly identical retrieval accuracy under exhaustive dense search, activation diversity represents a distinct structural property of embeddings that is uncaptured by traditional benchmark evaluation metrics.

\section{Conclusion}

This paper introduced the Hypergraph Embedding Index (HEI), a coordinate-inverted indexing framework that constructs inverted posting lists from higher-order combinations of latent embedding activations. By organizing documents through multi-dimensional hyperedges and combining complementary activation views, HEI achieves retrieval recall competitive with exhaustive vector search while pruning candidate evaluation pools. Furthermore, our analysis identified \emph{activation diversity} as a critical diagnostic metric governing coordinate indexability, establishing that uniform activation distributions are essential for efficient candidate generation in coordinate-inverted dense index structures.

\bibliographystyle{plain}

\end{document}